# Unidirectional Spin Hall Magnetoresistance in Antiferromagnetic Heterostructures

Yang Cheng[1,*], Junyu Tang[2], Justin J. Michel[3], Su Kong Chong[1], Fengyuan Yang[3], Ran Cheng[2], and Kang L. Wang[1,*]

[1]*Department of Electrical and Computer Engineering, and Department of Physics and Astronomy, University of California, Los Angeles, CA 90095, USA*

[2] *Department of Physics and Astronomy, University of California, Riverside, CA 92521, USA*

[3]*Department of Physics, The Ohio State University, Columbus, OH 43210, USA*

Corresponding author. E-mail: *cheng991@g.ucla.edu, *wang@ee.ucla.edu

Abstract

Unidirectional spin Hall magnetoresistance (USMR) has been widely reported in the heavy metal / ferromagnet (HM/FM) bilayer systems. We observe the USMR in the Pt/α-Fe$_2$O$_3$ bilayers where the α-Fe$_2$O$_3$ is an antiferromagnetic (AFM) insulator. Systematic field and temperature dependent measurements confirm the magnonic origin of the USMR. The appearance of AFM-USMR is driven by the imbalance of creation and annihilation of AFM magnons by spin orbit torque due to thermal random field. However, unlike its ferromagnetic counterpart, theoretical modeling reveals that the USMR in Pt/α-Fe$_2$O$_3$ is determined by the antiferromagtic magnon number, and with a non-monotonic field dependence. Our findings extend the generality of the USMR which pave the ways for the highly sensitive detection of AF spin state.

The detection of the spin state is one of the central topics in spintronics[1-3]. Spin Hall magnetoresistance (SMR) has been widely used to probe magnetization in heavy meatal/ (anti)ferromagnetic (HM/(A)FM) heterostructures[2,4-7]. When the current is applied in heavy metal layer, the generated spin current injects to the adjacent magnetic layer. The additional currents induced by inverse spin Hall effects change the resistivity of the heterostructures where its magnitude only depends on the relative angle between magnetization and spin polarization. So far, SMR has been widely reported in both ferromagnetic and antiferromagnetic systems. Recently, a new type of magnetoresistance --- unidirectional SMR (USMR)[8-11] --- has attracted intense interest. Compared with conventional SMR, the USMR is a non-linear magnetoresistance where the measured voltage depends quadratically on the applied current. What's more, the magnitude unidirectionally depends on the angle between spin polarization and magnetization as its name suggests, which provides a more precise way to probe the spin state in the magnetic layer. USMR has been observed only in HM/FM heterostructures. For metallic FM, the USMR is originated from either spin-dependent electron scattering (spin-dependent USMR) or electron-magnon scattering (spin-flip USMR) [9,12]. For insulating FM, for example Pt/$Y_3Fe_5O_{12}$ (YIG) bilayers[10], the observed USMR is attributed to the imbalance of magnon generation and annihilation rate by the spin-orbit torque. Although the SMR has been observed in both FM and AFM, the USMR is not expected in the AFMs. This is because the order parameter of AFM, the staggered magnetization (Néel vector), is a pair of two sublattice magnetizations, which makes the spin polarization with the 180º rotational symmetry relative to the Néel vector. A possible way to break this symmetry is by applying an external field to tilt the Néel vector. Here, we report the detection of USMR in Pt/α-$Fe_2O_3$ bilayers using the second harmonic measurement. Through systematic field and temperature dependent measurement, and theoretical modeling, we attribute the observed USMR to the imbalance of the magnon generation and annihilation rate, similar to the USMR in FM insulators [10]. However, the antiferromagnetic magnon plays a dominant role whereas the induced magnetization only contributes a small part to the USMR from our simulation.

α-Fe$_2$O$_3$ is an easy-plane antiferromagnet with the Néel temperature around 953 K [13]. The weak anisotropy field of three easy-axes in the *ab*-plane (0001) makes the spin-flop (or spin reorientation) occur at the critical field of ~ 1T, where the Néel order is perpendicular to the magnetic field, $\bm{n} \perp \bm{H}$ [14]. This makes the field control of the Néel order of α-Fe$_2$O$_3$ much easier compared with most of typical antiferromagnets, a necessary condition for the extraction of the USMR. We grow a Pt (5nm)/ α-Fe$_2$O$_3$ (30 nm) thin film stack on Al$_2$O$_3$ (0001) using off-axis magnetron sputtering [15]. Bulk α-Fe$_2$O$_3$ experiences the so-called Morin transition where it changes from easy-plane AFM to easy-axis (c-axis [0001]) AFM at ~ 260K. However, due to the strain induced enhancement of hard-axis anisotropy, our thin film α-Fe$_2$O$_3$ does not show such transition down to 10 K as had been demonstrated in our previous research[14-17]. After the growth of Pt/α-Fe$_2$O$_3$ bilayer, we pattern our sample into a Hall bar device with a length (l) 10 μm and a width (w) 5 μm, as shown in Fig. 1(a).

Since USMR is a non-linear current effect, an angular dependent harmonic measurement method is commonly used [10,18]. For the harmonic measurement setup, a low frequency 5 Hz ac current $I = I_0\sin(\omega t)$ is applied to the Hall bar device. The longitudinal ($V_{xx}$) and transverse ($V_{xy}$) voltage are measured simultaneously under in-plane magnetic field. The first harmonic response is the same as the DC measurement, where [19]

$$V_{xx}{}^{1\omega} = V_{\text{SMR}}\sin^2\varphi_{\text{H}} \tag{1}$$

$$V_{xy}{}^{1\omega} = -V_{\text{TSMR}}\sin\varphi_{\text{H}}\cos\varphi_{\text{H}}. \tag{2}$$

Here $\varphi_{\text{H}}$ is the in-plane angle between applied field and current direction, as shown in Fig 1(a). $V_{\text{SMR}}$ and $V_{\text{TSMR}}$ are the longitudinal and transverse spin Hall magnetoresistance where $\frac{V_{\text{SMR}}}{V_{\text{TSMR}}} = \frac{l}{w} = 2$ [14,20]. Figure 1(b) shows the angular dependent first harmonic measurement at 2 T and 300 K. The current $I_0 = 6$ mA. Figure 1(c) shows the extracted $V_{\text{SMR}}$ and $V_{\text{TSMR}}$ using Eq. (1) and (2) as the field increases from 0.3 T to 9 T. The magnitude of $V_{\text{SMR}}$ and $V_{\text{TSMR}}$ saturates near 1 T, which indicates single domain AFM state at $\mu_0 H > 1$ T. The ratio of $V_{\text{SMR}}$ and $V_{\text{TSMR}}$ is shown in the inset of Fig. 1(c), which is close to 2, as expected.

As shown in Fig. 2(a), a current induced effect such as spin-orbit torque drives the Néel order away from its equilibrium position where the change of magnetoresistance can be probed in the second harmonic voltage $V_{xx}^{2\omega}$ and $V_{xy}^{2\omega}$. Based on our previous research [19] [for more details, see Supplementary Materials (SM)], second harmonic voltage can be rewritten as

$$V_{xx}^{2\omega} = V_{xx,\text{FL}}^{2\omega} + V_{xx,\text{SSE}}^{2\omega} + V_{xx,\text{USMR}}^{2\omega}$$

$$= -\frac{1}{2}V_{\text{SMR}}\frac{H_{\text{FL}}}{H}\sin(2\varphi_H)\cos\varphi_H - (V_{xx,\text{SSE}} + V_{xx,\text{USMR}})\sin\varphi_H \quad (3)$$

$$V_{xy}^{2\omega} = V_{xy,\text{FL}}^{2\omega} + V_{xy,\text{SSE}}^{2\omega}$$

$$= \frac{1}{2}V_{\text{TSMR}}\frac{H_{\text{FL}}}{H}\cos(2\varphi_H)\cos\varphi_H + V_{xy,\text{SSE}}\cos\varphi_H. \quad (4)$$

Here, $H_{\text{FL}}$ is the field-like torque effective field. The longitudinal $V_{xx,\text{FL}}^{2\omega}$, $V_{xx,\text{SSE}}^{2\omega}$, and $V_{xx,\text{USMR}}^{2\omega}$ are the contributions from field-like torque, spin Seebeck effect, and USMR, respectively. The transverse $V_{xy,\text{FL}}^{2\omega}$ and $V_{xy,\text{SSE}}^{2\omega}$ are the field-like torque and spin Seebeck effect terms. Notice that USMR only shows up in the $V_{xx}^{2\omega}$ term. To extract the USMR contributions $V_{xx,\text{USMR}}$, we first fit the $V_{xx}^{2\omega}$ and $V_{xy}^{2\omega}$ to get the $V_{xx,\text{SSE}} + V_{xx,\text{USMR}}$ and $V_{xy,\text{SSE}}$; $V_{xx,\text{USMR}}$ can be separated from $V_{xx,\text{SSE}}$ given that $\frac{V_{xx,\text{SSE}}^{2\omega}}{V_{xy,\text{SSE}}^{2\omega}} = \frac{l}{w} = 2$. Compared with the fitting to ferromagnets, the transverse second harmonic voltage does not contain damping-like (DL) torque term, which makes the fitting more reliable. Figure 2 (b) and (c) shows the angular dependence of second harmonic voltage $V_{xx}^{2\omega}$ and $V_{xy}^{2\omega}$ at 2 T and 300 K. We fit the data using Eq. (3) and (4). Clearly, there is an USMR contribution in $V_{xx}^{2\omega}$ after subtracting the longitudinal spin Seebeck component $V_{xx,\text{SSE}}^{2\omega}$ with the same angular dependence.

Following the same process, we obtain $V_{xx}^{2\omega}$ and $V_{xy}^{2\omega}$ of the Pt(5 nm)/ α-Fe$_2$O$_3$ (30 nm) bilayer from 1 T to 9 T and fitting curves, as shown in Fig. 3(a) and (b). Figure 3 (c) shows the fitted field-like torque component from Fig. 3(a) and (b). The extracted $V_{xx,\text{FL}}^{2\omega}$ and $V_{xy,\text{FL}}^{2\omega}$ decrease with the increased field and follows a 1/H dependence. The ratio of $V_{xx,\text{FL}}^{2\omega}$ and $V_{xy,\text{FL}}^{2\omega}$ is shown in the blue curve of Fig. 3(e), which is around 2 for the entire field range as $\frac{V_{xx,\text{FL}}^{2\omega}}{V_{xy,\text{FL}}^{2\omega}} = \frac{V_{\text{SMR}}}{V_{\text{TSMR}}} = \frac{l}{w} = 2$. At the same time, the sinusoidal and cosinusoidal components in $V_{xx}^{2\omega}$ and $V_{xy}^{2\omega}$ are extracted by fitting the data with Eq. (3) and (4) where the ratio of them are plotted in the green curve of Fig. 3(e). The sinusoidal term in $V_{xx}^{2\omega}$ is expected to contain both longitudinal spin Seebeck and USMR contributions while the cosinusoidal term in $V_{xy}^{2\omega}$ is only from the transverse spin Seebeck voltage, which is linearly proportional to the field $H$ since $V_{\text{SSE}}^{2\omega} \propto \boldsymbol{m} \propto H$[21]. In the green curve of Fig. 3(e), it is shown that the ratio of $\frac{V_{xx,\text{SSE}}^{2\omega}+V_{xx,\text{USMR}}^{2\omega}}{V_{xy,\text{SSE}}^{2\omega}} > 2$, indicating the existence of USMR.

Figure 4(a) shows the field dependence of extracted USMR at 300 K. Surprisingly, unlike the USMR in a ferromagnet where the magnitude either monotonically decreases or is unchanged as the field increases, the USMR in the antiferromagnetic Pt/α-Fe$_2$O$_3$ bilayer shows a non-monotonic field dependence. The magnitude of USMR increases and reaches maximum at 2 T and then decreases and approaches zero. Since the AFM α-Fe$_2$O$_3$ is insulator, it excludes the possibility of spin-dependent or spin-flip mechanisms that require electron spin carriers. Recently, magnonic USMR has been observed in the insulating ferromagnetic bilayer Pt/YIG [10]. To testify the role played by magnons in the observed USMR, we perform the temperature dependent measurement as shown in Fig. 4(b). When the applied field is 2 T and the temperature decreases from 325 K, the USMR monotonically drops. At and below 200 K, no USMR is observed. The temperature dependence measurement provides the strong evidence for the magnonic origin of USMR [9].

Following the theory of magnon creation and annihilation imbalance in ferromagnet, we extend it to the antiferromagnetic regime. The coupled Landau–Lifshitz–Gilbert (LLG) equations are written as [22]

$$\dot{\boldsymbol{m}}_A = -\gamma \boldsymbol{m}_A \times \boldsymbol{H}_A^{eff} - \gamma J_{\mathrm{ex}} \boldsymbol{m}_A \times \boldsymbol{m}_B + \alpha \boldsymbol{m}_A \times \dot{\boldsymbol{m}}_A + \gamma \boldsymbol{\tau}_A^{\mathrm{DL}}, \tag{7}$$

$$\dot{\boldsymbol{m}}_B = -\gamma \boldsymbol{m}_B \times \boldsymbol{H}_B^{eff} - \gamma J_{\mathrm{ex}} \boldsymbol{m}_B \times \boldsymbol{m}_A + \alpha \boldsymbol{m}_B \times \dot{\boldsymbol{m}}_B + \gamma \boldsymbol{\tau}_B^{\mathrm{DL}}, \tag{8}$$

where the effective field $\boldsymbol{H}_{A(B)}^{eff} = \boldsymbol{H}_0 + \boldsymbol{h}_{A(B)} + \boldsymbol{H}_{A(B)}^{DMI} + \boldsymbol{H}_{A(B)}^{hard}$ contains the external magnetic field $\boldsymbol{H}_0$, thermal random field $\boldsymbol{h}_{A(B)}(T)$, effective field induced by Dzyaloshinskii–Moriya interaction (DMI) $\boldsymbol{H}_{A(B)}^{DMI} = H_D(\pm \boldsymbol{m}_{B(A)} \times \hat{\boldsymbol{z}})$ [23] and effective field of the hard axis anisotropy $\boldsymbol{H}_{A(B)}^{hard} = 2H_\perp m_{A(B)}^z \hat{\boldsymbol{z}}$. $J_{\mathrm{ex}}(< 0)$ is the AFM exchange coupling, $\gamma$ is the gyromagnetic ratio (positive), and $\alpha$ is Gilbert damping constant. $\boldsymbol{\tau}_{A(B)}^{\mathrm{DL}} = H_{\mathrm{DL}} \boldsymbol{m}_{A(B)} \times (\hat{\boldsymbol{\sigma}} \times \boldsymbol{m}_{A(B)})$ is the damping-like (DL) torque that exerts on the unit sublattice magnetization $\boldsymbol{m}_{A(B)}$. Here, $\hat{\boldsymbol{\sigma}}$ is the unit vector (along $\hat{\boldsymbol{y}}$ axis) of the spin polarization induced by spin Hall effect in Pt with amplitude $H_{\mathrm{DL}}$ being linearly proportional to the charge current density. In our previous work, we have demonstrated that without thermal random field, the DL torque induced the rotation of the sublattice magnetization (as well as the net magnetization $\boldsymbol{m}$ and Néel vector $\boldsymbol{n}$) $\Delta \varphi_{A(B)} = \Delta \varphi_m = \Delta \varphi_n \propto H_{\mathrm{DL}}^2 \propto I^2$ [19]. Therefore, the induced voltage change cannot be detected in second harmonic signal $V^{2\omega}$ but rather in third harmonic voltage $V^{3\omega}$, and it is not unidirectional. However, after considering the thermal random field $\boldsymbol{h}_{A(B)}(T)$, the damping-like torque induces fluctuation of sublattice magnetizations, which is now linear to the $I$, and unidirectional. The longitudinal SMR for a Pt/AFM insulator heterostructure can be characterized by $\rho_L = \rho_0 - \Delta \rho \langle n_y^2 \rangle$ [5,6,20], where the contribution from net magnetization $-\Delta \rho \langle m_y^2 \rangle$ is negligible in the AFM regime. With the thermal random field, the magnetic fluctuations of the sublattice magnetizations lead to the imbalance in the creation and annihilation of magnons, resulting in a USMR in $\rho_L$ [10]. From Eq.(3), the USMR signal reaches the maximum (minimum) at $\varphi_H = \pm \frac{\pi}{2} (\pm \pi)$. The USMR amplitude and antiferromagnetic magnon number difference are both proportional to the difference $\langle n_y^2 \rangle_+ - \langle n_y^2 \rangle_-$ ($\pm$ sign indicates $\varphi_H = \pm \frac{\pi}{2}$). In the following, we refer the term $\langle n_y^2 \rangle_+ - \langle n_y^2 \rangle_-$ as "antiferromagnetic magnon number difference" for convenience since the trivial proportionality does not affect the physical picture. To investigate the origin of USMR, we numerically calculate the field dependence of antiferromagnetic magnon number difference (see Section 1 in SM for more details). We find that the field dependence of antiferromagnetic magnon number difference qualitatively agrees with the non-monotonic trend of USMR signal and the peak around 2T [Fig. 4(a)] is reproduced [Fig. 5(a)]. We also calculate the antiferromagnetic magnon number difference between $\varphi_H = 0$ and $\varphi_H = \pi$ in Fig. 5(b). As expected, we find that $\langle n_y^2 (\varphi_H = 0) \rangle - \langle n_y^2 (\varphi_H = \pi) \rangle = 0$, which is consistent with the results from Fig. 2(a). The decrease of USMR at high fields is due to the suppression of magnon excitations at large $H_0$, which is similar to the Pt/FM case. However, unlike FMs where the magnetization saturates at small fields ($\boldsymbol{m} \parallel \boldsymbol{H_0}$), a small field will only cant of the sublattice magnetization of AFM ($\boldsymbol{n} \perp \boldsymbol{H_0}$), thus governing the magnetic fluctuation in a more complex way. Specifically, the canting angle $\Delta\varphi$ for sublattice magnetizations of $\alpha$-Fe$_2$O$_3$ depends on $H_D$, $J_{\text{ex}}$ and $H_0$ as $\Delta\varphi = \arcsin((H_0 + H_D)/-2J_{\text{ex}})$. With the thermal random field $\boldsymbol{h}$ acting on the two orthogonal $\hat{e}_\varphi$ and $\hat{e}_r$ directions, the dynamical magnetic susceptibility $\chi$ that characterizes the fluctuation of Néel vector $\Delta\boldsymbol{n}_y = \chi \cdot \boldsymbol{h}$ contains a highly nontrivial $H_0$ dependence through $\Delta\varphi$, leading to the non-monotonic field dependence of USMR in AFM. In Fig. S1(a) (See SM for details), we manually eliminate the $H_0$ dependence of $\Delta\varphi$ and plot the field dependence of magnon number difference. We find that the peak around 2 T disappears and the field dependence returns to be monotonic which is similar to the Pt/FM case. Therefore, the field-assist canting, which is unique in AFM, plays an important role for the non-monotonic field dependence of magnonic USMR. Finally, we compare the contributions with those of ferromagnetic magnon $\langle m_y^2 \rangle_+ - \langle m_y^2 \rangle_-$ as shown in Fig. S1(b), which is three order of magnitude smaller than that of antiferromagtic magnons, emphasizing the dominate role of antiferromagnetic magnons in the observed USMR.

In summary, we observe the USMR in the antiferromagnetic heterostructure in Pt/ α-Fe$_2$O$_3$ bilayers. The magnonic origin of USMR is revealed in the temperature and field dependence measurements. It is shown that the antiferromagnetic magnon plays the dominant role which gives a unique field dependence as compared with that of the ferromagnetic materials. This first evidence of USMR in HM/AFI bilayers significantly expands our materials base to include the large family of AF insulators and pave the ways for the highly sensitive detection of AF spin state in emerging the AF spintronics through USMR.

Note added: Recently we become aware of an independent report [24] which detected USMR in metallic AFM systems with similar nonmonotonic field dependence.

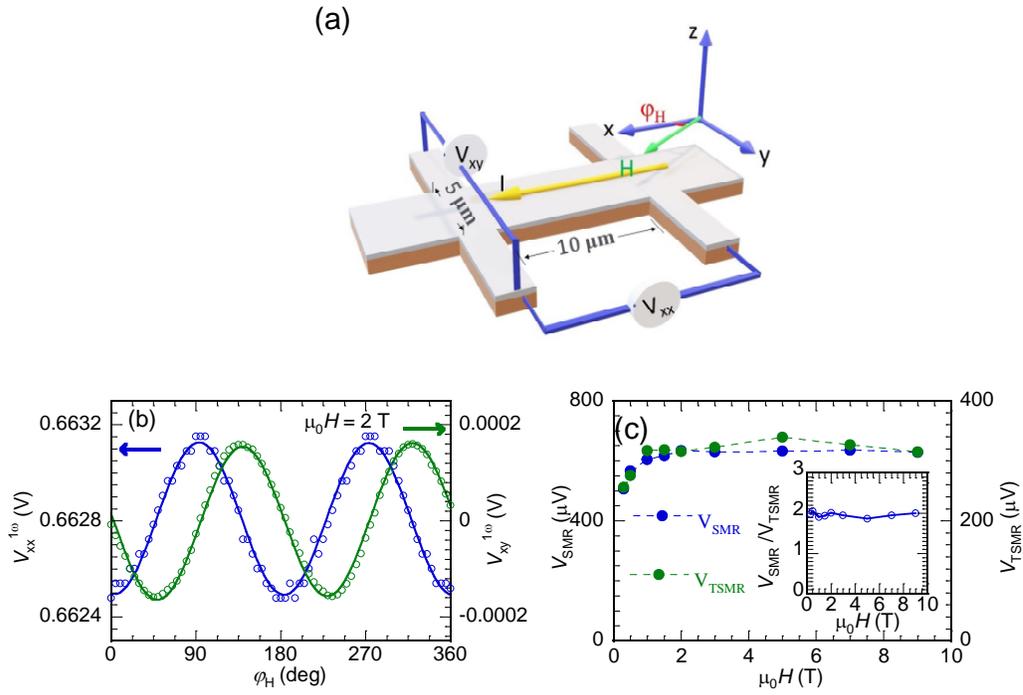

**Figure 1**. Experimental geometry and first harmonic results. (a) Schematics of a Pt/$\alpha$-Fe$_2$O$_3$ Hall bar with a 5 μm width and 10 μm length. (b) In plane angular dependence of first harmonic voltage $V_{xx}^{1\omega}$ (blue curve) and $V_{xy}^{1\omega}$ (green curve) for a Pt(5 nm)/$\alpha$-Fe$_2$O$_3$(30 nm) bilayer at 300 K with 2 T applied field. (c) Field dependence of (transverse) spin Hall magnetoresistance voltage $V_{\text{(T)SMR}}$ extracted from the fitting in (b) by Eq. (1) where the inset of (c) shows the ratio of $V_{\text{SMR}}$ and $V_{\text{TSMR}}$.

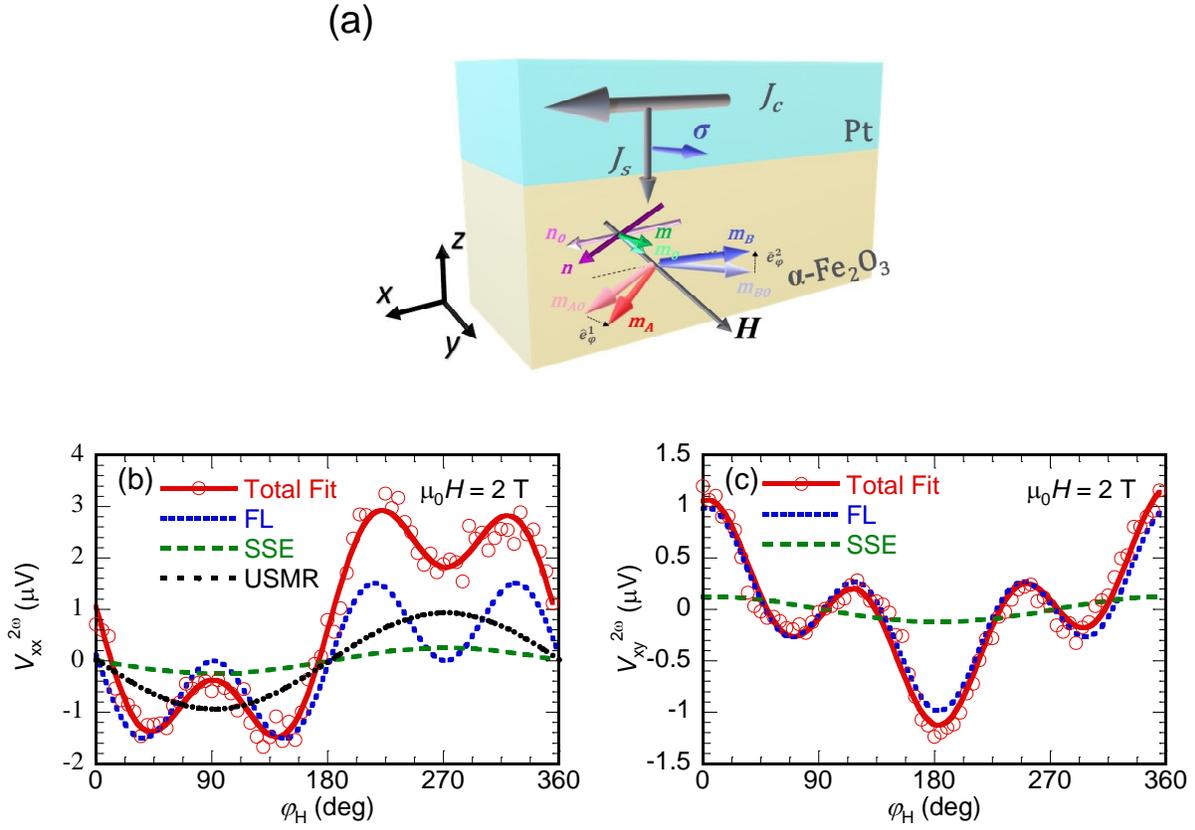

**Figure 2.** (a) Schematic of current induced spin orbit torque in two spin sublattices $m_{A(B)}$. In-plane angular dependence of second harmonic voltage (b) $V_{xx}^{2\omega}$ and (c) $V_{xy}^{2\omega}$ for a Pt(5 nm)/$\alpha$-Fe$_2$O$_3$(30 nm) bilayer at 300 K with 2 T applied field. The blue, green and black curves are contributions from the field-like torque, spin Seebeck effect, and USMR, respectively. The red curves are the total fit by Eq. (3) and (4).

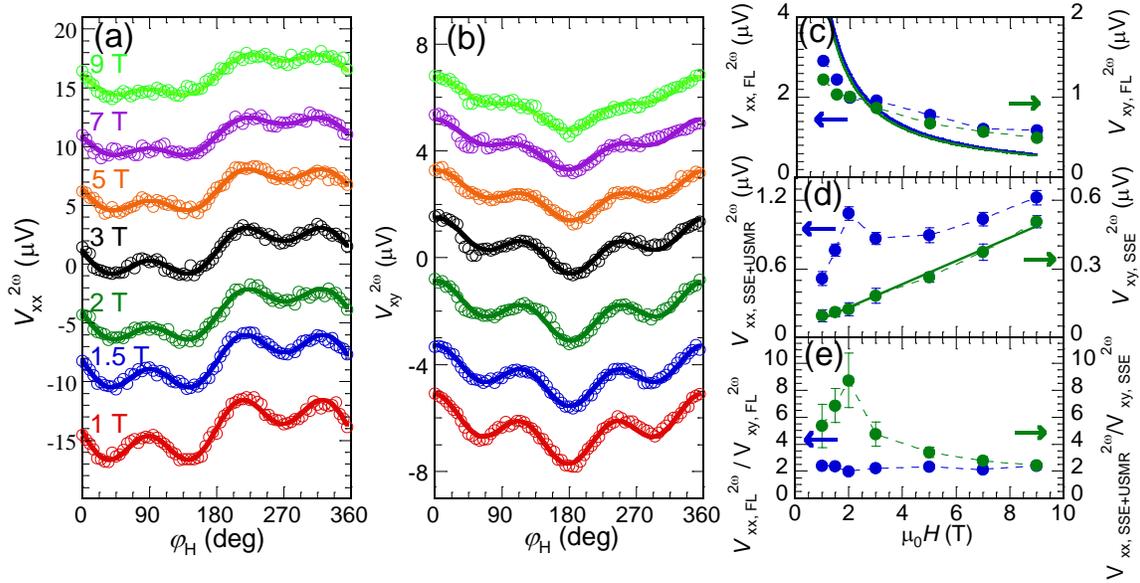

**Figure 3**. In-plane angular dependence of second harmonic Hall voltage (a) $V_{xx}^{2\omega}$ and (b) $V_{xy}^{2\omega}$ at different magnetic fields for a Pt(5 nm)/$\alpha$-Fe$_2$O$_3$(30 nm) bilayer at 300 K. (c) Field dependence of field-like torque contribution in $V_{xx}^{2\omega}$ (blue curve) and $V_{xy}^{2\omega}$ (green curve). The solid line is the 1/H fit. (d) Field dependence of spin Seebeck effect contribution in $V_{xx}^{2\omega}$ (blue curve) and $V_{xy}^{2\omega}$ (green curve). The solid line is the linear fit. (e) The ratio of $V_{xx,FL}$ and $V_{xy,FL}$ (blue curve) and the ratio of $V_{xx,\mathrm{SSE}} + V_{xx,\mathrm{USMR}}$ and $V_{xy,\mathrm{SSE}}$ (green curve), where the magnitude is calculated from (c) and (d). The ratio of $V_{xx,\mathrm{SSE}} + V_{xx,\mathrm{USMR}}$ and $V_{xy,\mathrm{SSE}}$ is greater than 2, which indicates the presence of USMR.

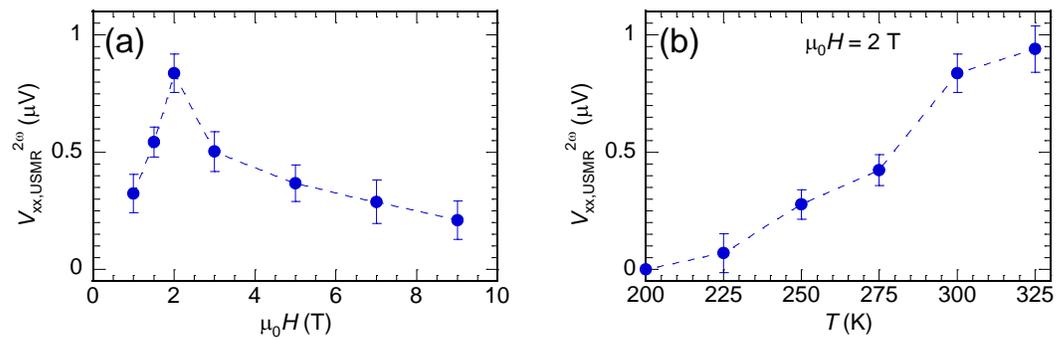

**Figure 4**. (a) The extracted magnetic field dependence of USMR contribution in the $V_{xx}^{2\omega}$. (b) Temperature dependence of USMR in $V_{xx}^{2\omega}$ at 2 T.

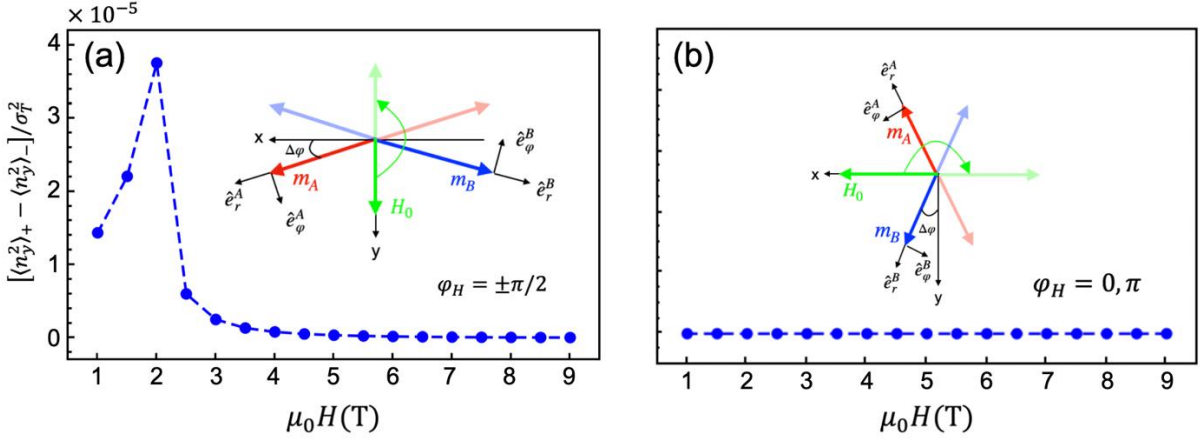

**Figure 5**. (a) Field dependence of antiferromagtic magnon number difference between $\varphi_H = \pm\pi/2$ and (b) between $\varphi_H = 0, \pi$. In the insets, light color arrows (red and blue) represent the magnetization after rotating $\boldsymbol{H_0}$ (green arrow) to the opposite direction. The magnetic fluctuation originates from the thermal random fields in the two orthogonal $\hat{e}_\varphi$ and $\hat{e}_r$ directions.